# SYSTEMIC CONSTRAINTS OF UNDECIDABILITY

Seth M. Bulin

University of Arizona, sethxbulin@gmail.com

ABSTRACT. This paper presents a theory of systemic undecidability, reframing incomputability as a structural property of systems rather than a localized feature of specific functions or problems. We define a notion of causal embedding and prove a closure principle: any subsystem that participates functionally in the computation of an undecidable system inherits its undecidability. This result positions undecidability as a pervasive constraint on prediction, modeling, and epistemic access in both natural and artificial systems. Our framework disarms oracle mimicry and challenges the view that computational limits can be circumvented through architectural innovation. By generalizing classical results into a dynamic systems context, this work augments the logical trajectory of Gödel, Turing, and Chaitin, offering a new perspective of the topology of computability and its interrelation to the boundaries of scientific knowledge.

## I. INTRODUCTION

Undecidability is traditionally framed at the level of specialized axiomatic systems and isolated functions[1-3]—originating with the halting problem and its lack of a uniform algorithmic solution. Modern systems continue to focus on isolated formulations of undecidability[4-11], without acknowledging that in predictive modeling, simulation, or emergent computation, undecidability rarely appears in isolation. Rather, it becomes embedded, entangled, and functionally necessary within larger architectures. This paper presents a theory of *systemic undecidability*: a formalism describing undecidability as a generalized structural constraint of complex systems. We define a notion of causal embedding to delineate the downward inheritance of undecidability from system to subsystem, and prove a novel closure principle situating the conditions under which this propagation occurs. Thus, undecidability is repositioned, not as a peripheral problem, but as a central axiom of systemic structure effectively limiting the computability of both total systems and their subcomponents.

By capturing the moment of functional entanglement between a subsystem's output with that of an undecidable system, we demonstrate that if such a subsystem is necessary for computing an undecidable function, then under minimal structural conditions, it too becomes undecidable. Thus, incomputability is not merely a local feature of isolated algorithms, but a property that spreads structurally across functional dependence. A visual model of a holographic system boundary illustrates this principle, where undecidability flows both inward and outward from an evolving terminus through causal links. From this perspective, complex systems are not simply at risk of undecidability, they are shaped by it. Our results extend classical limits of computation into a system-theoretic domain, with implications for both theoretical computer science and the modeling constraints of real-world intelligent systems[12,13].

This paper progresses as follows. First, we formalize systemic undecidability using a set-theoretic framework, and prove that undecidability propagates downward through functional or causal embedding into subsystems. Second, we apply this theory to show that even theoretically unconventional computer architectures, such as those claiming to simulate or approximate non-Turing-computable tasks, remain constrained by this principle. Third, we extend these implications into simulation theory, epistemology, and the predictive limits of scientific modeling. Finally, we position systemic undecidability in relation to the foundational contributions of Turing, Gödel, and Chaitin, distinguishing our theory as a structural generalization of classical limits. We conclude by considering whether undecidability is not merely a computational artifact, but a topological feature of reality itself.

## II. PRELIMINARY DEFINITIONS, NOTATIONS, AND CONCEPTIONS

We begin this section by defining key terms and introducing the conceptual foundations of computability theory—including closure properties, function composition, and system-level formalism—that underpin our results to follow. By briefly reviewing classical contributions, as found in both the standard literature[14,15,16] and original formulations[1,17], we establish the formal context in which our novel results and definitions are framed.

II.1 Classical Computability Notions

**Definition 2.1.** Let $\Sigma = \{0, 1\}$ be a finite binary alphabet, and let $\Sigma^*$ denote the set of all finite binary strings. We denote the set of natural numbers by $N$.

**Definition 2.2.** A *Turing machine M* is a tuple $(Q, \Sigma, \Gamma, \delta, q_0, q_{halt})$, where $Q$ is a finite set of states, $\Gamma \supseteq \Sigma$ is the tape alphabet, $\delta$ is the transition function, and $q_0, q_{halt}$ are the start and halting states respectively.

**Definition 2.3.** A function $f: \Sigma^* \to \Sigma^*$ is *computable* (or *recursive*) if there exists a Turing machine $M$ such that $M(x) = f(x)$ for all $x \in \Sigma^*$. We denote the set of all computable (total recursive) functions by REC. A function is *partially computable* if it is defined on a subset of inputs but may not halt on all $x \in \Sigma^*$.

II.2 Oracle Machines and Relative Computability

**Definition 2.4.** An *oracle Turing machine* is a Turing machine equipped with an oracle tape and query mechanism that allows it to decide membership in a language $O \subseteq \Sigma^*$ instantaneously. This enables *relative computability*: a function $f$ is computable relative to an oracle $O$ if a Turing machine with access to $O$ can compute $f$.

II.3 Closure Properties & Function Composition

**Definition 2.5.** Let $f, g: \Sigma^* \to \Sigma^*$ be total recursive functions. If $f, g \in$ REC, then their composition $f \circ g \in$ REC.

This closure under composition is a foundational result. However, in what follows, we do not limit ourselves to isolated function composition, but rather study how undecidability propagates structurally across systems and subsystems in a computational universe.

II.4 Systems and Computational Universes

**Definition 2.6.** We define a system $A$ as a structured subset of a universe $U$, where $U$ is the space of discrete-state computational processes over binary inputs. Each system $A \subseteq U$ is associated with a (possibly partial) function $f_A : \Sigma^* \to \{0, 1\} \cup \{\uparrow\}$, representing its observable behavior. The output $f_A(x)$ denotes the system's response to input $x$, where $\uparrow$ indicates nontermination.

**Definition 2.7.** A subsystem $B \subseteq A$ is said to be *causally embedded* in $A$ if the computation of $f_A(x)$ depends functionally on the output of $f_B(x)$ over some domain $D \subseteq \Sigma^*$. That is, there exist computable functions $\psi$ and $z$ such that:

$$f_A(x) = z(f_B(x), \psi(x)) \text{ for all } x \in D$$

This formulation captures *functional necessity*: the behavior of $f_B$ is nontrivially entangled with the computation $f_A$, and cannot be abstracted away without loss of functionality.

Our notion of causal embedding allows us to model how computational constraints—undecidability—can propagate downward from a system to its integrants. In particular, if $f_A$ is undecidable and $f_B$ is causally embedded in $A$, then $f_B$ inherits that undecidability under suitable closure conditions, as formalized in the next section.

Further formal clarifications on functional necessity, computational universes, and sufficient domains are provided in *Appendix A*.

## III. A FORMAL THEORY OF SYSTEMIC UNDECIDABILITY

We now introduce a formalism that reconceptualizes undecidability as an innate structural principle of systems—rather than a property confined to the solvability of individual functions or decision problems[1,13,14]. Classical computability theory treats undecidability as localized, for example, the halting problem is formally undecidable because no Turing machine can solve it on all inputs. But this limitation is traditionally tethered to the behavior of specific algorithms, formal languages, or axiomatic systems[1-3].

In contrast, we generalize undecidability as a property that can propagate structurally within a system composed of functionally dependent components. Specifically, we examine when the undecidability of a global system imposes constraints on its subsystems via structural entanglement; from this, a closure principle for systemic undecidability arises, viz. any subsystem causally embedded in an undecidable system inherits that undecidability under minimal conditions.

Note that this result is not a reformulation of classical closure under composition[15]. In recursion theory, if functions $f$ and $g$ are computable, then their composition $f \circ g$ is also computable. Our setting differs: we are not analyzing operations on computable functions, but rather how the undecidability of a system constrains the computability of its embedded systems. The following definition formalizes this relationship.

III.1 Causal Embedding & Functional Dependence

**Definition 3.1.** Let $A \subseteq U$ be a discrete state-system whose global behavior is governed by a function $f_A : \Sigma^* \to \{0, 1\} \cup \{\uparrow\}$, where $\Sigma^*$ is the set of finite binary strings. We say that $A$ is undecidable if $f_A \notin$ REC.

III.2 Structural Closure Theorem

We now formalize our main result: if a system $A$ is undecidable, and a subsystem $B \subseteq A$ is causally embedded in $A$ over a sufficient domain, then $B$ is also undecidable. This establishes a systemic constraint that extends beyond function-level undecidability, reframing incomputability as a structurally entangled property.

III.3 Structural Closure of Undecidability

**Theorem 3.1.** Let $A \subseteq U$ be a system with output function $f_A : \Sigma^* \to \{0, 1\} \cup \{\uparrow\}$, and suppose $A$ is undecidable (i.e., $f_A \notin \text{REC}$). Let $B \subseteq A$ be a subsystem that is causally embedded in $A$ over a domain $D \subseteq \Sigma^*$, such that there exist total computable functions $\psi$ and $z$ satisfying:

$$f_A(x) = z(f_B(x), \psi(x)) \text{ for all } x \in D$$

If $D$ is sufficient to determine the undecidability of $f_A$ (e.g., $f_A|_D \notin \text{REC}$), then $f_B \notin \text{REC}$. That is, $B$ is also undecidable.

This result does not rely on a particular syntactic coding of functions or machines. Rather, it reveals a *structural property of systems that express computation*: if an undecidable function depends essentially on the behavior of a subsystem, then that subsystem cannot be computable in general, lest it collapse the undecidability of the global system.

**Sketch of Proof.** Suppose, for contradiction, that $f_B \in \text{REC}$. Since both $\psi$ and $z$ are total computable, and $f_B$ is assumed computable, the composition $x \mapsto z(f_B(x), \psi(x))$ would be computable. But by assumption, this function is invariant to $f_A(x)$ for all $x \in D$, and $f_A|_D \notin \text{REC}$. This contradicts the closure of REC under composition. Therefore, $f_B \notin \text{REC}$.

III.4 Interpretation

This result establishes a *structural closure property* for undecidability. The principal idea is that undecidability propagates structurally through causal embedding: if a subsystem's output is functionally necessary for computing the behavior of an undecidable system, then that subsystem itself must be undecidable. Thus, the implication is not merely logical, but architectural, viz. no subsystem of an undecidable system can resolve the system's undecidability if it participates causally in that computation. This formally grounds the idea that systemic undecidability imposes global limits on inference, simulation, and architectural abstraction.

III.5 Visualizing Structural Propagation

To clarify the intuition behind Theorem 3.1, consider a system $A$ whose computational boundaries are visualized as a holographic plane in which subsystems are embedded. The undecidability of A is not confined to a local function or edge case but instead permeates the causal structure of the entire system. When a subsystem $B \subseteq A$ is causally embedded, viz. its behavior is necessary for computing $f_A$, undecidability bidirectionally emanates through that dependency. The result is topological propagation of incomputability from an undecidable boundary encoding: undecidability becomes a structurally inherited property and an expansive constraint, not merely a singular anomaly or local feature.

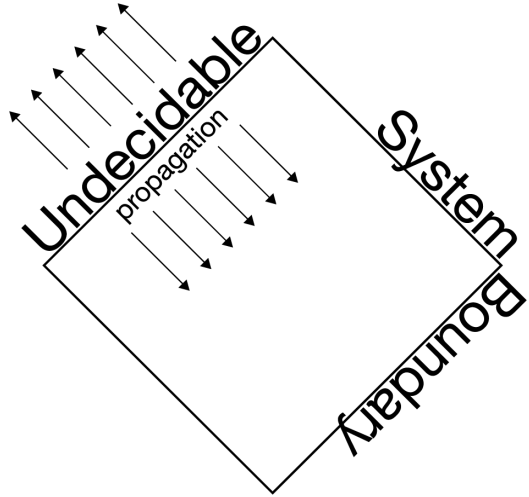

**Figure 1.** A conceptual diagram illustrating the propagation of undecidability within a structured computational system. The holographic boundary encoding represents the undecidable system *A*, while the interior regions constitute causally embedded subsystems $B_1, B_2,\ldots$, whose outputs are functionally entangled with $f_A$. Arrows indicate bidirectional causal embedding across a computational domain. According to Theorem 3.1, each subsystem that is functionally necessary over a sufficient domain inherits the undecidability of *A*. This models the structural closure of undecidability across causal architecture, that is, undecidability becomes *structurally intrinsic to the entire causal topology of the system*.

## IV. THE ORACLE FALLACY & UNCONVENTIONAL ARCHITECTURES

Attempts to overcome undecidability often invoke Turing's notion of an oracle—a metamathematical machine capable of solving problems beyond the means of effective computation[1,17]. Such a conception, as formalized in Section II.2, empowers speculative models to try to simulate oracle behavior within physical or networked systems[18]. By mimicking oracular output behavior across finite instances, it is hoped that these models might circumvent classical computing limits.

One such approach[4] suggests that sufficiently large, interconnected networks can "learn the undecidable" by leveraging small-diameter topologies and an imitative synergistic protocol for the fittest nodes. Nodes in such a system adopt the behavior of the most successful peers—those producing higher outputs—in an attempt to decide instances of the halting problem with high probability. As the problem space grows, the system is claimed to approach asymptotic solvability.

Other models appeal to hypercomputation[5-10], proposing that exotic architectures—those involving closed timelike curves[6], relativistic computation[7,10], or analog infinities[8,9]—might access solutions beyond the Turing limit. Yet these approaches fail to acknowledge a deeper constraint: any attempt to reify a true oracle within a simulated or physical architecture remains bounded by the computability of the system itself. Following the closure principle established in Section III, any such architecture inherits the undecidability of the system it attempts to model or contain.

Positing invariance between a simulated or mimicked oracle and a true oracle constitutes a fallacy. A true oracle yields correct answers to undecidable problems with perfect generality and no time or resource bounds. Simulations, however emergent or probabilistically reliable, operate within Turing-bounded constraints and thus cannot resolve undecidability itself.

These restrictions extend beyond computer architecture and into epistemology by constraining what simulation can predict, know, or explain. Thus, systemic undecidability reframes not only the limits of computability but the limits of inference itself, placing boundaries on algorithmic reasoning and scientific prediction[19]. In the next section, we magnify the consequences of this boundary for modeling and inference in complex systems.

## V. UNDECIDABLE IMPLICATIONS FOR MODELING & EPISTEMOLOGY

Effects of the structural closure principle of undecidability reach beyond the theory of computation and grasp the domains of modeling, simulation, and epistemology. If the undecidability of a system structurally constrains all functionally entangled subsystems, as Theorem 3.1 formalizes, then no subsystem-based model can fully resolve or approximate the behavior of the global system without itself inheriting that computational intractability.

Such a result directly limits simulation-based inference. Common scientific practice involves modeling complex systems by isolating subsystems and simulating their behavior under emergent or idealized rules[20,21]. However, if these subsystems participate in undecidable dynamics[22,23], then the simulations are bounded[24] not merely by measurement error or empirical unknowns, but by *logical inaccessibility*; as a result, probabilistic models and emergent behaviors are impacted: even if system-level patterns stabilize, their foundations remain structurally incomputable.

These constraints recontextualize our understanding of epistemic access[19,25,26]. We cannot, in general, approximate or isolate the behavior of undecidable systems[27,28] by modeling their subsystems, no matter how adaptive or opulent the modeling architecture may be[29,30]. Thus, a *nec plus ultra* is placed not just on the tools of science, but on the structure of experience[31] and knowledge itself [32], viz. what can be known is bounded by the structural undecidability of the systems in which we are embedded.

Our result is not a dismissal of the scientific armamentarium, but a realization: modeling becomes an exercise in situated inference, constrained by incompleteness, and simulation becomes a heuristic[33]—powerful, but epistemically limited. In this view, systemic undecidability becomes the horizon of scientific generalization and the logical foreground of modeling humility. In the next and final section we position our contributions alongside the works of Gödel, Turing, and Chaitin.

## VI. BEYOND GÖDEL, TURING, AND CHAITIN: SYSTEMIC UNDECIDABILITY AS STRUCTURE

The theory of systemic undecidability developed in this paper builds upon, and augments, the foundations built by Gödel[2], Turing[1,17], and Chaitin[3]. Each of these thinkers revealed incompleteness or incomputability as intrinsic features of formal systems and algorithmic processes; but they framed these limitations as local or formal, confined to specific statements, theorems, or functions.

Gödel's incompleteness theorems showed that no sufficiently expressible formal system can prove all truths about its own composition. Turing's analysis of the halting problem revealed the limits of mechanical computation. And Chaitin further extended these ideas into algorithmic information theory, identifying randomness as a fundamental limit to algorithmic compression.

Each work shares a focus on the bounds of specific deductive or computational acts. What this paper contributes is a generalization of undecidability, that is, undecidability is not merely local—it propagates structurally. The closure principle proved in Section III formalizes the idea that when a system is undecidable, its functionally entangled subsystems cannot evade this limitation. Thus, undecidability becomes a property of architecture, not just output.

Gödel, Turing, and Chaitin initiated a trajectory that has now been extended into the architecture of systems themselves. The move from pointwise constraints to systemic structure repositions undecidability as a foundational principle not only in mathematical-logic or computer science, but across physics, epistemology, modeling, and inference. In doing so, it invites a shift in how we conceive the limits of our own conscious inference and our attempts to simulate any part of a system that is structurally undecidable. If those earlier insights marked the boundaries of formalism and algorithmic reasoning, the theory presented here supplies us with the topology of the map itself, leaving what we model, simulate, and inhabit fundamentally unchartable.

## VII. CONCLUSION

This paper has introduced a general theory of systemic undecidability, formalizing the notion that undecidability is not confined to individual problems or isolated functions, but can propagate structurally through a system's internal architecture. Through the definition of causal embedding and the proof of a closure principle, we have shown that functionally necessary subsystems of an undecidable system inherent its incomputability. This extends the classical results of Gödel, Turing, and Chaitin into a new framework—one in which undecidability emerges not as a local anomaly, but as a systemic and global constraint. As increasingly complex artificial and analytical systems are developed and simulated, this work outlines a principle of *metaundecidability*: a structural limit on what can be computed, modeled, or epistemically accessed within computationally entangled domains.

## REFERENCES

1. Turing, A. M. "On computable numbers, with an application to the Entscheidungsproblem." *Proc. Lond. Math. Soc.* **42**, 230–265 (1936).

2. Gödel, K. Über formal unentscheidbare Sätze der Principia Mathematica und verwandter Systeme I. *Monatsh. Math. Phys.* **38**, 173–198 (1931).

3. Chaitin, G. J. Information-theoretic limitations of formal systems. *J. ACM* **21**, 403–424 (1974).

4. Abrahão, F. S. et al. *Learning the undecidable from networked systems. arXiv* preprint arXiv:1904.07027v3 [cs.DC] (2021).

5. López-Díaz, A. J., Pelayo, F. J., Montero, F. & Mira, J. The origin and evolution of information handling. *arXiv* preprint arXiv:2404.04374v5 [physics.bio-ph] (2024).

6. Andréka, H., Németi, I. & Szekely, G. "Closed timelike curves in relativistic computation." *Nat. Comput.* **11**, 1–10 (2012).

7. Andréka, H., Németi, I. & Németi, P. "General relativistic hypercomputing and foundation of mathematics." *Nat. Comput.* **8**, 499–516 (2009).

8. Aaronson, S. "The limits of quantum computers." *Sci. Am.* **298**, 62–69 (2008).

9. Deutsch, D. Quantum theory, the Church–Turing principle and the universal quantum computer. *Proc. R. Soc. Lond. A* **400**, 97–117 (1985).

10. Earman, J. & Norton, J. D. Forever is a day: Supertasks in Pitowsky and Malament–Hogarth spacetimes. *Philos. Sci.* **60**, 22–42 (1993).

11. Stepney, S. et al. *Journeys in Non-Classical Computation* (Springer, 2008).

12. Giere, R. How models are used to represent reality. *Philos. Sci.* **71**, 742–752 (2004).

13. Mitchell, M. *Complexity: A Guided Tour* (Oxford Univ. Press, 2009).

14. Sipser, M. *Introduction to the Theory of Computation* (3rd ed.). Cengage Learning (2012).

15. Arora, S. & Barak, B. *Computational Complexity: A Modern Approach*. Cambridge University Press (2009).

16. Rogers, H. Jr. *Theory of Recursive Functions and Effective Computability*. MIT Press (1987).

17. Turing, A. M. "Systems of logic based on ordinals." *Proc. Lond. Math. Soc.* **45**, 161–228 (1939).

18. Copeland, B. J. & Proudfoot, D. "Alan Turing's forgotten ideas in computer science." *Sci. Am.* **280**, 76–81 (1999).

19. Humphreys, P. Extending ourselves: Computational science, empiricism, and scientific method. *Oxford Univ. Press* (2004).

20. Rosen, R. *Life Itself: A Comprehensive Inquiry into the Nature, Origin, and Fabrication of Life. Columbia Univ. Press* (1991).

21. Giere, R. How models are used to represent reality. *Philos. Sci.* **71**, 742–752 (2004).

22. Wegner, P. Why interaction is more powerful than algorithms. *Commun. ACM* **40**, 80–91 (1997).

23. Crutchfield, J. P. Between order and chaos. *Nat. Phys.* **8**, 17–24 (2011).

24. Piccinini, G. Computational modeling vs. mechanistic explanation. *Synthese* **154**, 1–20 (2007).

25. Frigg, R. & Reiss, J. The philosophy of simulation: Hot new issues or same old stew? *Synthese* **169**, 593–613 (2009).

26. Floridi, L. *The Philosophy of Information* (Oxford Univ. Press, 2011).

27. Colombo, M. & Seriès, P. Bayes in the brain—On Bayesian modelling in neuroscience. *Brit. J. Philos. Sci.* **63**, 697–723 (2012).

28. Bostrom, N. "Are you living in a computer simulation?" *Philos. Q.* **53**, 243–255 (2003).

29. Leonelli, S. Data-centric biology: A philosophical study. *Univ. Chicago Press* (2016).

30. Berkeley, I. S. N. "The curious case of connectionism." *Behav. Brain Sci.* **11**, 64–65 (1988).

31. Chalmers, D. J. "Facing up to the problem of consciousness." *J. Conscious. Stud.* **2**, 200–219 (1995).

32. Turing, A. M. "Computing machinery and intelligence." *Mind* **59**, 433–460 (1950).

33. Winsberg, E. Simulated experiments: Methodology for a virtual world. *Philos. Sci.* **70**, 105–125 (2003).

## APPENDIX A. FORMAL CLARIFICATIONS

A.1 Functional Necessity

We say that a subsystem $B \subseteq A$ is functionally necessary for computing $f_A$ over domain $D \subseteq \Sigma^*$ if there exist total computable functions $\psi$ and $z$ such that:

$$f_A(x) = z(f_B(x), \psi(x)) \text{ for all } x \in D$$

and there exists no alternative computable function $g$ such that $f_A(x) = g(\psi(x))$ for all $x \in D$. That is, $f_B$ cannot be abstracted away without altering the behavior of $f_A$ on that domain. This notion captures structural entanglement, in contrast to logical sufficiency or equivalence.

A.2 Universe of Discrete-State Systems

The computational universe $U$ is defined as the class of all discrete-state systems $A$ such that each system maps inputs $x \in \Sigma^*$ to inputs $\{0, 1\} \cup \{\uparrow\}$, where $\uparrow$ denotes non-termination. These systems may include Turing machines, distributed processes, or networked architectures, as long as their behavior can be described over binary sequences.

A subsystem $B \subseteq A$ is any functionally dependent component whose output behavior $f_B$ contributes to $f_A$ under a computable embedding. We remain agnostic to internal state encoding or architectural specifics, focusing instead on the observable dependency between $f_A$ and $f_B$.

A.3 Sufficient Domains

A domain $D \subseteq \Sigma^*$ is said to be sufficient for undecidability propagation if $f_A|_D$ is undecidable, that is, no computable function can replicate $f_A(x)$ for all $x \in D$. This ensures that the contribution of $f_B$ to $f_A$ on $D$ cannot be attributed to a trivial or otherwise decidable fragment.

The notion of sufficiency here is not topological but logical, that is, the undecidability of $f_A$ is essentially supported by the behavior over $D$. In this context, structural closure of undecidability follows as a consequence of causal embedding over sufficient domains.